\documentclass{aa}
\usepackage{graphics}

\begin{document}

   \thesaurus{01     
              (02.02.1;  
               13.07.3;  
               13.07.1;  
               13.07.2)}  
   \title{On the Pair Electromagnetic Pulse of a Black Hole with
Electromagnetic Structure}


   \author{Remo Ruffini
          \inst{1}
          \and
          Jay D. Salmonson\inst{2}
	    \and
	    James R. Wilson\inst{2}
	    \and          
	    She-Sheng Xue\inst{1}
          }

   \institute{I.C.R.A.-International Center for Relativistic Astrophysics
and
Physics Department, University of Rome ``La Sapienza", I-00185 Rome,
Italy\\
	\and Lawrence Livemore National Laboratory, University of
     California, Livermore, California, U.S.A.
             }


   \maketitle

   \begin{abstract}

We study the relativistically expanding electron-positron pair plasma
formed by the process of vacuum polarization around an electromagnetic
black hole (EMBH). Such processes can occur for EMBH's with mass all
the way up to $6\cdot 10^5M_\odot$.  Beginning with a idealized model
of a Reissner-Nordstrom EMBH with charge to mass ratio $\xi=0.1$, 
numerical hydrodynamic calculations are
made to model the expansion of the pair-electromagnetic pulse (PEM
pulse) to the point that the system is transparent to photons. Three idealized 
special relativistic models have been compared and contrasted with the results 
of the numerically integrated general relativistic hydrodynamic equations.
One of the three models has been validated: a PEM pulse of constant thickness 
in the laboratory frame is shown to be in excellent agreement with results 
of the general relativistic hydrodynamic code. It is remarkable that this 
precise model, starting from the fundamental parameters of the EMBH, leads 
uniquely to the explicit evaluation of the parameters of the PEM pulse, including 
the energy spectrum and the astrophysically unprecedented large Lorentz factors (up to $6\cdot 10^3$ for a $10^3 M_{\odot}$
EMBH). The observed photon energy at the peak
of the photon spectrum at the moment of photon decoupling is shown to
range from 0.1 MeV to 4 MeV as a function of the EMBH
mass. Correspondingly the total energy in photons is in the range of
$10^{52}$ to $10^{54}$ ergs, consistent with observed gamma-ray bursts. 
 In these computations we neglect the presence of baryonic matter
which will be the subject of forthcoming publications.

   \end{abstract}

%

\section{\it Introduction}

At present there is a need for viable source models capable of
producing the large energy ($\sim 10^{53} - 10^{54}$ ergs) necessary to
power distant gamma-ray bursts such as GRB971214 (e.g.~\cite{fp}) and
GRB990123 (\cite{bloom}).  In this paper we present a model for the
production of such large bursts based on results of calculations of
the energy emission from vacuum polarization processes around an EMBH.

Damour \& Ruffini (1975) indicated a very precise mechanism to explain the origin
of gamma ray bursts: the vacuum polarization of a Kerr-Newman EMBH 
 via the Heisenberg-Euler-Schwinger process. They showed how these processes, 
occurring for EMBH's all the way up to $6\cdot 10^5M_\odot$, can radiate a large
fraction of the extractable energy of the EMBH, in the sense of the Black Holes
reversible transformations (\cite{rc}). Recently, after the discovery of the 
gamma ray afterglow by the Beppo Sax satellite and their optical identification (e.g.~\cite{fp}),
 this model has been reconsidered by Ruffini (1998) who introduced the new 
concept of the ''dyadosphere" of an EMBH and by Preparata et al. (1998a,1998b) who 
have given details of the spatial density and energy distribution in the 
electron-positron plasma created in the dyadosphere.
The present paper examines the time evolution of such an electron-positron 
plasma by assuming an EMBH in  vacuum and calculates the consequences.

Past work on black hole formation by Wilson (1975, 1977) has
shown that highly charged EMBH's with charge to mass ratio $\xi=Q /(M\sqrt{G})\approx 0.1$, surrounded by 
an oppositely charged magnetosphere, 
may form during the collapse of rotating magnetized stars.  For a low
mass star to collapse and form a highly charged black hole, surrounded 
by an oppositely charged magnetosphere, strong
initial fields and rapid rotation are required. The theoretical background 
for such an analysis  has been presented in Ruffini \& Wilson (1975), and 
Damour et al. (1978). The recent discovery
of the soft gamma-ray repeater SGR1806-20 (\cite{sgr1,sgr2}) indicates
that very high magnetic fields may be present in precollapse stars of
a few solar masses. 

The calculations by Wilson 
assumed space was filled with a perfect conductor so that the
magneto-hydrodynamic model could be used.  In the region above a
newly formed EMBH the density is falling rapidly.  Eventually,
due to the paucity of charge carriers, the assumption of the plasma as
a perfect conductor must fail, possibly leading to regions where the
electric fields will be sufficiently high so that pair creation will
ensue. Similar scenarios can be envisaged for gravitational collapse 
of much larger masses $M\sim 10^4-10^5 M_\odot$ occurring in galactic nuclei.
In the present paper we begin with the idealized
situation of an isolated, Reissner-Nordstrom EMBH with charge to mass ratio $\xi=0.1$. The case of a Ker Newmann EMBH is presented in \cite{tdrr}. The vacuum polarization 
processes are computed, following the approach outlined in Ruffini (1998), 
Preparata et al. (1998), Ruffini (1999a), Ruffini et al. (1999), Ruffini (1999b), 
and they give the specific initial data of the dyadosphere.
The evolution of
the PEM pulse formed by the vacuum polarization processes is followed by a numerical
calculation to the point where the system is transparent to photons.
We then analyze the final gamma-ray signal that would be observed from
this idealized model of an EMBH in  vacuum.  Extensions of
this model such as the expansion of the PEM pulse in the presence of
the remnant baryonic matter left over in the collapse process will be the subject of a forthcoming publication\cite{rssw1}. Some consequences for the interpretation of the observed burst and afterglow are presented in \cite{pr1} and \cite{prx3}.

\section{\it Formation of $e^+e^-$ pairs from an EMBH}

We recall the relevant formulas from Ruffini(1998) and Preparata et al. 
(1998a, 1998b): 

The Christodoulou-Ruffini mass
formula for a Reissner-Nordstrom black hole  gives (\cite{rc})
\begin{eqnarray}
E&=&Mc^2=M_{\rm ir}c^2 + {Q^2\over2r_+},\label{em}\\
S&=& 4\pi r_+^2=16\pi\left({G^2\over c^4}\right) M^2_{\rm ir},
\label{sa}
\end{eqnarray}
with
\begin{equation}
{\sqrt{G}Q\over r_+c^2}\leq 1,
\label{s1}
\end{equation}
where $M_{\rm ir}$ is the irreducible mass, $r_{+}=2M_{\rm ir}$ is the
horizon radius, $S$ is the horizon surface area, and maximally charged
($Q_{\rm max}=2\sqrt{G}M_{ir}$) black holes satisfy the equality in
Eq.(\ref{s1}).

The dyadosphere is defined as the region outside the horizon of a
EMBH where the electric field exceeds the
critical value for spontaneous $e^+e^-$ pair creation, ${\cal E}_{\rm
c}={m^2c^3\over\hbar e}$ (\cite{he}, \cite{sw}), where $m$ and $e$ are the mass
and charge of the electron.  By introducing the dimensionless mass
and charge parameters $\mu={M\over M_{\odot}}$, $\xi={Q\over Q_{\rm
max}}\le 1$ for Reissner-Nordstrom black holes, the horizon radius may be
expressed as
\begin{eqnarray}
r_{+}&=&{GM\over c^2}\left[1+\sqrt{1-{Q^2\over GM^2}}\right]\nonumber\\
&=&1.47 \times 10^5\mu (1+\sqrt{1-\xi^2})\hskip0.1cm {\rm cm}.
\label{r+}
\end{eqnarray}
The outer limit of the dyadosphere is
defined as the radius $r_{\rm ds}$ at which the electric field produced by the
EMBH equals the critical field. The radius of the dyadosphere can be 
expressed in terms of the
Planck charge $q_{\rm p}=(\hbar c)^{1\over2}$ and the Planck mass
$m_{\rm p}=({\hbar c\over G})^{1\over2}$  in the form
\begin{eqnarray}
r_{\rm ds}&=&\left({\hbar\over mc}\right)^{1\over2}\left({GM\over
c^2}\right)^{1\over2} \left({m_{\rm p}\over m}\right)^{1\over2}\left({e\over
q_{\rm p}}\right)^{1\over2}\left({Q\over \sqrt{G}M}\right)^{1\over2}\nonumber\\
&=&1.12\cdot 10^8\sqrt{\mu\xi} \hskip0.1cm {\rm cm},
\label{rc}
\end{eqnarray} 
which well illustrates the hybrid gravitational and quantum nature of
this quantity. The radial interval $r_{+}\leq r \leq r_{\rm ds}$ describes the region 
in which $e^+e^-$ pairs are produced (see Fig.[1] of \cite{prxa}).

In a very short time $\sim O({\hbar\over mc^2})$, a very large number
of pairs is created in the dyadosphere.  The local number density of
pairs created is computed as a function of radius
\begin{equation}
n_{e^+e^-}(r) = {Q\over 4\pi r^2\left({\hbar\over
mc}\right)e}\left[1-\left({r\over r_{ds}}\right)^2\right] ~,
\label{nd}
\end{equation}
and the energy density is given by
\begin{equation}
\epsilon(r) =  {Q^2 \over 8 \pi r^4} \biggl(1 - \biggl({r \over
r_{ds}}\biggr)^4\biggr) ~, \label{jayet}
\end{equation}
(see Figs.[2] \& [3] of \cite{prxa}). 
The total energy of pairs converted from the static electric energy
and deposited within the dyadosphere is then
\begin{equation}
E^{\rm tot}_{e^+e^-}={1\over2}{Q^2\over r_+}(1-{r_+\over r_{\rm ds}})\left[1-
\left({r_+\over r_{\rm ds}}\right)^2\right] ~.
\label{tee}
\end{equation}

In the limit of ${r_+\over r_{ds}}\rightarrow 0$, Eq.(\ref{tee}) 
leads to 
$E^{\rm tot}_{e^+e^-}\rightarrow {1\over2}{Q^2\over r_+}$, which
coincides with the energy 
extractable from EMBH's by reversible processes ($M_{ir}={\rm const.}$), 
namely $E-M_{ir}={1\over2}{Q^2\over r_+}$(see Fig.[4] of \cite{prxa}).

Due to the very large pair density given by Eq.(\ref{nd}) and to
the sizes of the cross-sections for the process $e^+e^-\leftrightarrow
\gamma+\gamma$, the system is expected to thermalize to a plasma
configuration for which
\begin{equation}
N_{e^+}=N_{e^-}=N_{\gamma}=N_{\rm p}
\label{plasma}
\end{equation}
and reach an average temperature
\begin{equation}
kT_\circ={ E^{\rm tot}_{e^+e^-}\over3N_{\rm p}\cdot2.7},
\label{t}
\end{equation}
where $N_p=N_{\rm pair}$ is the number of $e^+e^-$-pairs created in the 
dyadosphere and $k$ is Boltzmann's constant
(see Fig.[4] of Preparata et al. 1998b). 

In conclusion, we stress that (i) the locality of $e^+e^-$ pair
creation is an important feature of the dyadosphere, and (ii) $e^+e^-$ pair
creation occurs only in a finite range of the parameters $\mu$ and $\xi$ and 
a unique spectrum of the created pairs can be calculated in terms of these two parameters $\mu,\xi$
(iii) Eqs.(\ref{em}-\ref{tee}) show that the pair-creation
process is capable of extracting large amounts of energy from an EMBH
with an extremely high efficiency (close to $100\%$).

\begin{figure}
   \resizebox{\hsize}{8cm}{\includegraphics{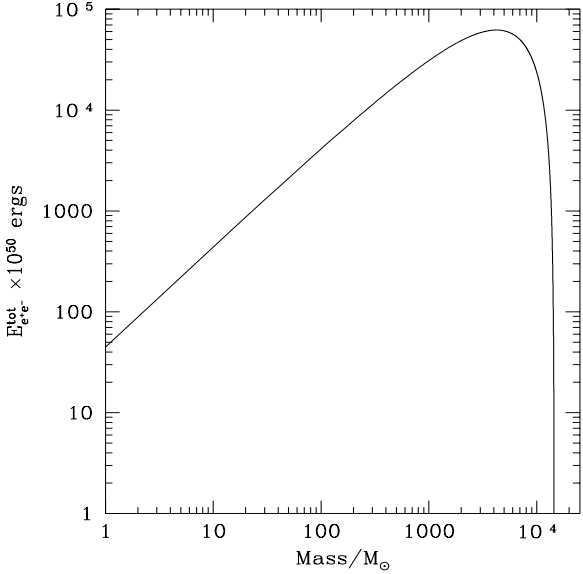}}
      \caption[]{ The total energy in $e^+e^-$ pairs deposited
within the dyadosphere of a black hole with mass $M$ and charge 
$Q = 0.1 Q_{\rm max}$ equal
to $10 \%$ of maximal black hole charge.
\label{figetot}}
   \end{figure}

\begin{figure}
   \resizebox{\hsize}{8cm}{\includegraphics{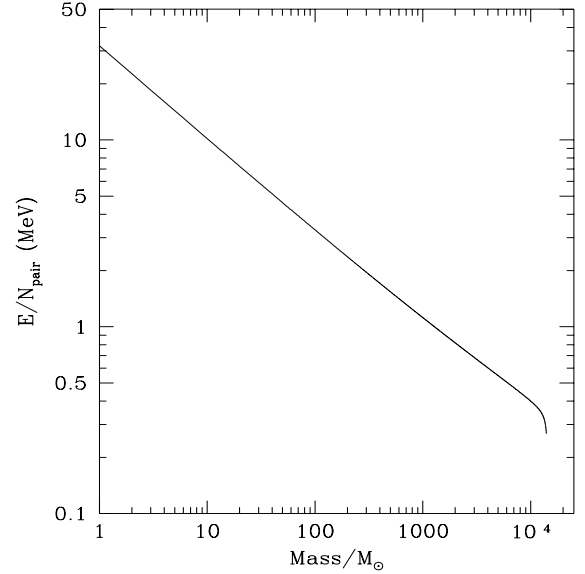}}
      \caption[]{ An average energy per pair  versus
mass is given for a black hole with mass $M$ and charge equal
to $10 \%$ of maximal black hole charge: $\xi=0.1$.
\label{fig71}}
   \end{figure}

In Wilson (1975, 1977) a black hole charge of the order $Q = 0.1
Q_{\rm max}(\xi = 0.1)$ was formed. Thus, we henceforth assume an EMBH
with $\xi = 0.1$ for our detailed numerical calculations.  
In Fig.[\ref{figetot}] a plot of the total energy Eq.(\ref{tee}) versus mass
is given for the case $\xi = 0.1$ and in Fig.[\ref{fig71}] a plot of
the average energy per pair (relating to Eq.(\ref{t})) versus mass is
given also for $\xi = 0.1$.
%

%

\section{\it General relativistic hydrodynamic equations}

In order to model the radial evolution of the $e^+e^-$-pair and photon plasma fluid created
in the dyadosphere of an EMBH, we need to discuss the
relativistic hydrodynamic equations governing such an evolution and the associated rate equation for the pairs.

The metric for a Reissner-Nordstrom black hole is
\begin{equation}
ds^2=-g_{tt}(r)dt^2+g_{rr}(r)dr^2+r^2d\theta^2 +r^2\sin^2\theta
d\phi^2 ~,
\label{sw}
\end{equation}
where $g_{tt}(r)= - \left[1-{2GM\over c^2r}+{Q^2G\over
c^4r^2}\right] \equiv - \alpha(r)^2$ and $g_{rr}(r)= \alpha(r)^{-2}$.

The plasma fluid of $e^+e^-$-pairs, photons and baryon matter in the curved spacetime given by Eq.(\ref{sw}) is described by the covariant stress-energy tensor given by (\cite{ws})
\begin{equation}
T^{\mu\nu}= pg^{\mu\nu}+(p+\rho)U^\mu U^\nu + \Delta T^{\mu\nu},
\label{tensor'}
\end{equation}
where $\rho$ and $p$ are respectively the total proper energy density and pressure in the comoving frame. $U^\mu$ is the four-velocity
of the plasma fluid. $\Delta T^{\mu\nu}$ accounts for dissipative effects due to heat conduction, shear and bulk viscosity. In this paper, as a first approximation, we assume
the plasma fluid of $e^+e^-$ pairs, photons and baryon matter to be a simple perfect
fluid in curved spacetime and neglect the dissipative terms $\Delta T^{\mu\nu}$ in Eq.(\ref{tensor'}). In spherical symmetry we have
\begin{equation}
g_{tt}(U^t)^2+g_{rr}(U^r)^2=-1 ~,
\label{tt}
\end{equation}
where $U^r$ and $U^t$ are the radial and temporal contravariant
components of the 4-velocity.  The conservation law for the  baryon number
can be expressed as a function of the proper baryon number density
$n_B$ as follows
\begin{eqnarray}
(n_B U^\mu)_{;\mu}&=& g^{-{1\over2}}(g^{1\over2}n_B
U^\nu)_{,\nu}\nonumber\\
&=&(n_BU^t)_{,t}+{1\over r^2}(r^2 n_BU^r)_{,r}=0 ~.
\label{contin}
\end{eqnarray}
The energy-momentum conservation law of the plasma fluid is
\begin{eqnarray}
(T_\mu^\sigma)_{;\sigma}&=&{\partial p\over\partial x^\mu}+
g^{-{1\over2}}{\partial \over\partial
x^\nu}\left[g^{1\over2}(p+\rho)U_\mu U^\nu\right]\nonumber\\
& +& {1\over2}{\partial
g_{\nu\lambda}\over\partial x^\mu} (p+\rho)U^\nu U^\lambda=0 ~.
\label{conse}
\end{eqnarray}
The radial component reduces to 
\begin{eqnarray}
{\partial p\over\partial r}&+&{\partial \over\partial t}\left[(p+\rho)U^t U_r\right]
+{1\over r^2} { \partial
\over \partial r}  \left[r^2(p+\rho)U^r U_r\right]\nonumber\\
&-&{1\over2}(p+\rho)\left[{\partial g_{tt}
 \over\partial r}(U^t)^2+{\partial g_{rr}
 \over\partial r}(U^r)^2\right] =0 ~.
\label{cmom2}
\end{eqnarray}
The component of the energy-momentum conservation equation
(\ref{conse}) along a flow line is
\begin{eqnarray}
U_\mu(T^{\mu\nu})_{;\nu}&=&-(\rho U^\nu)_{;\nu}
-pU^\nu_{;\nu}\nonumber\\ &=&-g^{-{1\over2}}(g^{1\over2}\rho
U^\nu)_{,\nu} - pg^{-{1\over2}}(g^{1\over2} U^\nu)_{,\nu}\nonumber\\
&=&(\rho U^t)_{,t}+{1\over r^2}(r^2\rho
U^r)_{,r}\nonumber\\
&+& p\left[(U^t)_{,t}+{1\over r^2}(r^2U^r)_{,r}\right]=0 ~.
\label{conse1}
\end{eqnarray}
Defining the total proper internal energy density $\epsilon$ and the baryon mass density $\rho_B$ in the comoving frame
\begin{equation}
\epsilon \equiv \rho - \rho_B,\hskip0.5cm \rho_B\equiv n_Bm_Bc^2 ~,
\label{cpp}
\end{equation} 
and using the law of baryon-number conservation (\ref{contin}), from
Eq.(\ref{conse1}) we have
\begin{equation}
(\epsilon U^\nu)_{;\nu} +pU^\nu_{;\nu}=0 ~.
\label{conse'}
\end{equation}
Recalling that ${dV\over d\tau}=VU^\mu_{;\mu}$, where $V$ is the
comoving volume and $\tau$ is the proper time for the fluid, we have
along each flow line
\begin{equation}
{d(V\epsilon)\over d\tau}+p{dV\over d\tau}={dE\over d\tau}+p{dV\over
d\tau}=0 ~,
\label{f'}
\end{equation}
where $E=V\epsilon$ is total proper internal energy of the plasma fluid.

Finally, we represent the equation of state by a thermal index
$\Gamma(\rho,T)$
\begin{equation}
\Gamma = 1 + { p\over \epsilon} ~.
\label{state}
\end{equation}

We now turn to the analysis of $e^+e^-$ pairs initially created in
the dyadosphere. Let $n_{e^-}$ and $n_{e^+}$ be the proper densities of
electrons and positrons.  We have
\begin{equation}
n_{e^-}=n_{e^+}=n_{\rm p} ~.
\label{eee}
\end{equation}
The rate equation for electrons and positrons gives
\begin{eqnarray}
(n_{e^\pm}U^\mu)_{;\mu}&=&(n_{e^\pm}U^t)_{,t}+{1\over r^2}(r^2 n_{e^\pm}U^r)_{,r}\nonumber\\
&=&\overline{\sigma v} \left[n_{e^-}(T)n_{e^+}(T) - n_{e^-}n_{e^+}\right] ~,
\label{econtin}
\end{eqnarray}
where $\sigma$ is the mean pair annihilation-creation cross-section,
$v$ is the thermal velocity of $e^\pm$, $\overline{\sigma v}$ is the average value of $\sigma v$ and $n_{e^\pm}(T)$
are the proper number-densities of electrons and positrons that thermalize with
photons through the process $e^+e^-\leftrightarrow
\gamma+\gamma$, $n_{e^\pm}(T)=n_{\gamma}(T)$ for $T>m_ec^2$, where $n_{\gamma}(T)$ 
is the number-density of photons. The proper number-densities $n_{e^\pm}(T)$
are given by
appropriate Fermi integrals with zero chemical potential, at the equilibrium temperature $T$. The equilibrium temperature
$T$ is determined by the thermalization processes occurring in 
the PEM-pulse with a total proper energy-density $\rho$, whose evolution is
governed by the hydrodynamic Eqs.(\ref{contin},\ref{cmom2},\ref{conse1}). We have
\begin{equation}
\rho = \rho_\gamma + \rho_{e^+}+\rho_{e^-}+\rho^b_{e^-}+\rho_B,
\label{eeq}
\end{equation}
where $\rho_\gamma=a T^4$ is the photon energy-density and $\rho^b_{e^-}$
is the energy-density of the electrons stemming from the ionization
of baryonic matter, which is given by a Fermi integral with an
appropriate chemical potential $\mu_e$ at the equilibrium temperature
$T$, and $\rho_B\simeq m_Bc^2n_B$, where $n_B$ is the baryon
number-density, since baryons and ions
are expected to be non-relativistic in the range of temperature $T$
under consideration. $\rho_{e^\pm}$ is the
proper energy-density of electrons and positrons of pairs defined by
\begin{equation}
\rho_{e^\pm} \equiv {n_{e^\pm}\over n_{e^\pm}(T)}\rho_{e^\pm}(T),
\label{hat}
\end{equation}
where $n_{e^\pm}$ is obtained by integration of Eq.(\ref{econtin}) and 
$\rho_{e^\pm}(T)$ is the proper energy-density of
electrons (positrons) obtained from zero chemical potential Fermi
integrals at the equilibrium temperature $T$. Having implicitly
defined the equilibrium temperature $T$ by Eq.(\ref{eeq}), we
can also analogously evaluate the total pressure
\begin{equation}
p = p_\gamma + p_{e^+}+p_{e^-}+p^b_{e^-}+p_B,
\label{eep}
\end{equation}
where $p_\gamma$ is the photon pressure, $p_{e^\pm}$ defined by
\begin{equation}
p_{e^\pm} \equiv {n_{e^\pm}\over n_{e^\pm}(T)}p_{e^\pm}(T),
\label{hat'}
\end{equation}
where the pressures $p_{e^\pm}(T)$ are determined by zero chemical
potential Fermi-integrals at the equilibrium temperature $T$, and
$p^b_{e^-}$ is the pressure of the ionized electrons, evaluated by an
appropriate Fermi integral of non-zero chemical potential $\mu_e$. In
Eq.(\ref{eep}), the ion pressure $p_B$ is negligible by comparison
with the pressures $p_{\gamma, e^\pm, e^-_b}(T)$. Finally, using Eqs.(\ref{eeq},\ref{eep}), we
compute the thermal index $\Gamma$ of the equation of state
(\ref{state}).

It is clear that in order to be correct, the entire set of equations
considered above, namely Eqs.(\ref{contin},\ref{cmom2},\ref{conse1}) 
with equation of state given
by Eq.(\ref{state}) and the rate equation (\ref{econtin}) have
to be integrated fulfilling the total energy conservation for the
system.
The boundary conditions adopted here are simply purely ingoing conditions at the horizon 
and outgoing at radial infinity.

The calculation is initiated by depositing a proper energy density
(\ref{jayet}) between the Reissner-Nordstrom horizon radius $r_+$ and the
dyadosphere radius $r_{ds}$.  The total energy deposited is given by
Eq.(\ref{tee}). The calculation is continued as the plasma fluid expands,
cools and the $e^+e^-$ pairs recombine, until it becomes optically
thin:
\begin{equation} 
\int_R dr(n_{e^\pm}+\bar
Zn_B)\sigma_T\simeq O(1),
\label{thin}
\end{equation}
where $\sigma_T =0.665\times 10^{-24}
{\rm cm^2}$ is the Thomson cross-section is , $\bar Z$ is the average atomic number of baryon matter
and the integration is over the radial interval of the PEM-pulse in the
comoving frame. At this point the energy is virtually entirely in the
form of free-streaming photons and the calculation is stopped. In the following computations we shall assume that baryonic matter is absent. The case of presence of baryonic matter, left over in a remnant during the process of gravitational collapse, is discussed in \cite{rssw1}.

\section{\it Numerical integration of general relativistic hydrodynamic equations }

We use a computer code (Wilson et al. 1997, 1998) to
evolve the spherically symmetric hydrodynamic equations for the
baryons, $e^+e^-$-pairs and photons deposited in the dyadosphere.
  
We define the generalized Lorentz factor $\gamma$ and the radial
coordinate velocity $V^r$
\begin{equation}
\gamma \equiv \sqrt{ 1 + U^r U_r},\hskip0.5cm V^r\equiv {U^r\over U^t}.
\label{asww}
\end{equation}
From Eqs.(\ref{sw}, \ref{tt}), we then have
\begin{equation}
(U^t)^2=-{1\over g_{tt}}(1+g_{rr}(U^r)^2)={1\over\alpha^2}\gamma^2.
\label{rr}
\end{equation}
Following Eq.(\ref{cpp}), we also define 
\begin{equation}
E \equiv \epsilon \gamma,\hskip0.5cmD \equiv \rho_B \gamma,
\hskip0.3cm {\rm and}\hskip0.3cm\tilde\rho \equiv \rho\gamma
\label{cp}
\end{equation} 
so that the conservation law of baryon number (\ref{contin}) can then
be written as
\begin{equation}
{\partial D \over \partial t} = - {\alpha \over r^2} {
\partial \over \partial r} ({r^2 \over \alpha} D V^r).
\label{jay1}
\end{equation}
Consequently, Eq.(\ref{conse1}) acquires the form,
\begin{equation}
{\partial E \over \partial t} = - {\alpha \over r^2} {
\partial \over \partial r} ({r^2 \over \alpha} E V^r) - p
\biggl[ {\partial \gamma \over \partial t} + {\alpha \over r^2}
{\partial \over \partial r} ({ r^2 \over \alpha} \gamma V^r)
\biggr].
\label{jay2}
\end{equation}
Defining the radial coordinate momentum density
\begin{equation}
S_r\equiv \alpha (p+\rho)U^tU_r = (D + \Gamma E) U_r,  
\label{mstate}
\end{equation}
we can express the radial component of the energy-momentum
conservation law given in Eq.(\ref{cmom2}) by
\begin{eqnarray}
{\partial S_r \over \partial t} &=& - {\alpha \over r^2} { \partial
\over \partial r} ({r^2 \over \alpha} S_r V^r) - \alpha {\partial p
\over \partial r}\nonumber\\ 
&-&{\alpha\over2}(p+\rho)\left[{\partial g_{tt}
\over\partial r}(U^t)^2+{\partial g_{rr} \over\partial
r}(U^r)^2\right]\nonumber\\ 
&=& - {\alpha \over r^2} { \partial \over
\partial r} ({r^2 \over \alpha} S_r V^r) - \alpha {\partial p \over
\partial r}\nonumber\\
&-& \alpha\left({M \over r^2}-{Q^2 \over r^3}\right)
\biggl({D + \Gamma E \over \gamma} \biggr) \biggl[ \left({\gamma \over
\alpha}\right)^2 + {(U^r)^2 \over \alpha^4 } \biggr] ~.
\label{jay3}
\end{eqnarray}

In order to determine the number-density of $e^+e^-$ pairs, we turn to Eq.(\ref{econtin}). 
Defining the $e^+e^-$-pair coordinate density
$N_{e^\pm} \equiv\gamma n_{e^\pm}$ and $N_{e^\pm}(T) \equiv\gamma
n_{e^\pm}(T)$ and using Eq.(\ref{rr}), we rewrite the rate equation given by Eq.(\ref{econtin}) in the form
\begin{equation}
{\partial N_{e^\pm} \over \partial t} = - {\alpha \over r^2} {
\partial \over \partial r} ({r^2 \over \alpha} N_{e^\pm} V^r) +
\overline{\sigma v} (N^2_{e^\pm} (T) - N^2_{e^\pm})/\gamma^2~,
\label{jay:E:ndiff}
\end{equation}
where the equilibrium temperature $T$ has been obtained from
Eqs.(\ref{eeq}) and (\ref{hat}).

\section{\it A simple special relativistic treatment of the expansion 
of the PEM-pulse}
 
In the following, using the definitions introduced in Section 4, we
study the general properties of PEM-pulses. As the zeroth order
approximation, we do not consider: (i) the gravitational interaction
since it can be considered a perturbation against the total energy of
the PEM-pulse and we describe the expanding fluid by a
special relativistic set of equations, (ii) the radiation energy lost
during expansion since this is a very small fraction of total energy
before the dense PEM-pulse becomes transparent.

Analogous to Eq.(\ref{f'}), from Eq.(\ref{contin}), we
have, in the general case of presence of baryonic matter, along each flow line,
\begin{equation}
{d(n_BV)\over d\tau}=0.
\label{f0}
\end{equation}
For a small shell of volume $\Delta V$ at radius $r$, for the expansion
of that shell from the volume $\Delta V_\circ$ to the volume $\Delta  V$,
we obtain
\begin{equation}
{n_B^\circ\over n_B}= {\Delta V\over \Delta V_\circ}={\Delta {\cal V}\gamma(r)
\over \Delta {\cal V}_\circ\gamma_\circ(r)},
\label{be'}
\end{equation}
where $\Delta {\cal V}$ is the volume of the shell in the coordinate
frame and it relates to the proper volume $\Delta V$ in the comoving
frame by $\Delta V=\gamma(r) \Delta {\cal V}$, where $\gamma(r)$
defined in Eq.(\ref{asww}) is the $\gamma$-factor of the shell
at the radius $r$.

Similarly from Eq.(\ref{f'}), using the equation of state
(\ref{state}),  along the flow lines we obtain
\begin{equation}
d\ln\epsilon + \Gamma d\ln V=0.
\label{scale''}
\end{equation}
With a very small variation of the volume from $\Delta V_\circ$ to $\Delta V$ 
and the internal energy density $\epsilon$ along the flow lines we obtain 
\begin{equation}
{\epsilon_\circ\over \epsilon} = 
\left({\Delta V\over \Delta V_\circ}\right)^\Gamma=
\left({\Delta {\cal V}\over \Delta {\cal V}_\circ}\right)^\Gamma\left({\gamma(r)
\over \gamma_\circ(r)}\right)^\Gamma ~,
\label{scale'}
\end{equation}
where the thermal index $\Gamma$ given by (\ref{state}) is computed in for each value of $\epsilon,p$ as a function of $\Delta V$.  The thermal index $\Gamma$ is a
slowly-varying function of the state with values around $4/3$.

We know that as the PEM-pulse expands, the overall energy conservation
requires that the change of total internal proper energy of a shell in the
fluid will be compensated by a change in the bulk kinetic energy of
the shell. We clearly have for the total internal proper energy of 
the PEM-pulse in the comoving frame
\begin{equation}
{\cal E}_{\Delta {\cal V}}=\int_{\Delta V}\rho dV=\int_{\Delta {\cal V}}
\tilde\rho d{\cal V},
\end{equation}
where $dV(d{\cal V})$ is the proper (coordinate) volume of a fluid
element in the shell of the PEM-pulse, and $\tilde\rho$ is defined by
(\ref{cp}).  The change of the kinetic energy of a fluid element in
the shell of the PEM-pulse due to an infinitesimal expansion is given by,
\begin{equation}
dK=[\gamma(r)-1](dE+\rho_BdV).
\label{dk}
\end{equation}
The total energy conservation for that shell implies
\begin{equation}
{\cal E}_{\Delta {\cal V}_\circ}-{\cal E}_{\Delta {\cal V}}=\int_{\Delta {\cal V}_\circ}\tilde\rho_\circ d{\cal V}-\int_{\Delta {\cal V}}\tilde\rho d{\cal V}
=\int_{\Delta {\cal V}_\circ}^{\Delta {\cal V}}dK,
\label{k}
\end{equation}
where ${\cal E}_{\Delta {\cal V}}$ is the total internal proper energy of 
the shell including baryon masses.

In order to model the relativistic expansion of plasma
fluid analytically, we assume that $E$ and $D$, as defined by Eq.(\ref{cp}), are
constant in space over the volume $\Delta V$.  As a consequence, we can describe the system in terms of the relativistic Lorentz factor
$\gamma(r)$, the volume $\Delta {\cal V}$ in coordinate frame and the
quantities $\epsilon$ and $\rho_B$ known in the comoving frame. We can
rewrite the left-hand-side of Eq.(\ref{k}) as
\begin{equation}
(\epsilon_\circ+\rho^\circ_B)\gamma_\circ (r) {\Delta \cal V}_\circ
-(\epsilon+\rho_B)\gamma(r) {\Delta \cal V},
\label{lhs}
\end{equation}
and the right-hand-side of Eq.(\ref{k}) as
\begin{equation}
(\epsilon+\rho_B){\Delta \cal V}\gamma(r) (\gamma(r)-1)
-(\epsilon_\circ+\rho^\circ_B){\Delta \cal V}_\circ\gamma_\circ (r) (\gamma_\circ (r)-1).
\label{rhs}
\end{equation}
Thus Eq.(\ref{k}) simplifies to
\begin{equation}
(\epsilon_\circ+\rho^\circ_B)\gamma_\circ^2(r) {\Delta \cal V}_\circ =(\epsilon+\rho_B)
\gamma^2 (r){\Delta \cal V},
\label{res'}
\end{equation}
which leads the solution
\begin{equation}
\gamma (r)=\gamma_\circ(r)\sqrt{{(\epsilon_\circ+\rho^\circ_B){\Delta \cal V}_\circ
\over(\epsilon+\rho_B) {\Delta \cal V}}}.
\label{result'}
\end{equation}

In the special relativistic case the rate equation (\ref{econtin}) for $e^\pm$ reduces to
\begin{eqnarray}
{\partial (n_{e^\pm}\gamma(r)) \over \partial t} &=& - {1\over r^2} {
\partial \over \partial r} (r^2n_{e^\pm}\gamma(r) V^r)\nonumber\\
 &+&
\overline{\sigma v} (n^2_{e^\pm} (T) - n^2_{e^\pm}).
\label{special}
\end{eqnarray}
In the shell of the PEM-pulse, we define the coordinate
number-density of $e^\pm$ in equilibrium to be
$N_{e^\pm}(T)\equiv\gamma(r) n_{e^\pm}(T)$ and correspondingly the coordinate
number-density of $e^\pm$ to be $ N_{e^\pm}\equiv\gamma(r) n_{e^\pm}$.
With the assumption that $n_{e^\pm}$ in Eq.(\ref{econtin}) is constant in space over
the volume $\Delta V$, we obtain the equation for the evolution of the $e^\pm$
coordinate number-density as they fall out of equilibrium with the
PEM-pulse, as seen by a coordinate observer
\begin{equation}
{\partial \over \partial t}(N_{e^\pm}) = -N_{e^\pm}{1\over\Delta {\cal V}}{\partial \Delta {\cal V}\over \partial t}+\overline{\sigma v}{1\over\gamma^2(r)}  (N^2_{e^\pm} (T) - N^2_{e^\pm})~.
\label{paira'}
\end{equation}
Eqs.(\ref{be'}), (\ref{scale'}), (\ref{result'}) and
(\ref{paira'}) are a complete set of equations describing the
relativistic expansion of the shell in the PEM-pulse.

If we now turn from a single shell of the PEM-pulse to a finite
distribution of shells over the PEM-pulse, we introduce the average values of the proper
internal-energy, baryon-mass, baryon-number and pair-number densities ($\bar\epsilon,
\bar\rho_B,\bar n_B,\bar n_{e^\pm}$), and $\bar E\equiv\bar\gamma\bar\epsilon$,
$\bar D\equiv\bar\gamma\bar\rho_B$, $\bar N_{e^\pm}\equiv\bar\gamma(r) \bar n_{e^\pm}$ for the PEM-pulse, where the average relativistic
$\bar\gamma$-factor is defined by,
\begin{equation}
\bar\gamma={1\over{\cal V}}\int_{\cal V}\gamma(r) d{\cal V},
\label{ga}
\end{equation}
and ${\cal V}$ is the total coordinate volume of the PEM-pulse. 
Considering the volume expansion of the PEM-pulse from ${\cal V}_\circ$ to ${\cal V}$, we
transform Eqs.(\ref{be'},\ref{scale'}) into
\begin{eqnarray}
{\bar n_B^\circ\over \bar n_B}&=& { V\over  V_\circ}={ {\cal V}\bar\gamma
\over {\cal V}_\circ\bar\gamma_\circ},
\label{be}\\
{\bar\epsilon_\circ\over \bar\epsilon} &=& 
\left({V\over V_\circ}\right)^\Gamma=
\left({ {\cal V}\over  {\cal V}_\circ}\right)^\Gamma\left({\bar\gamma
\over \bar\gamma_\circ}\right)^\Gamma,
\label{scale}
\end{eqnarray}
where $ V=\bar\gamma {\cal V}$ is the total proper volume of the PEM-pulse, following
from the definition $\bar\gamma$ in Eq.(\ref{ga}).  Analogously,
Eq.(\ref{res'}) becomes
\begin{equation}
(\bar\epsilon_\circ+\bar\rho^\circ_B)\bar\gamma_\circ^2 {\cal V}_\circ =(\bar\epsilon+\bar\rho_B){\cal V}\bar\gamma^2,
\label{res}
\end{equation}
and Eq.(\ref{result'}) becomes
\begin{equation}
\bar\gamma =\bar\gamma_\circ\sqrt{{(\bar\epsilon_\circ+\bar\rho^\circ_B){\cal V}_\circ
\over(\bar\epsilon+\bar\rho_B) {\cal V}}},
\label{result}
\end{equation}
while Eq.(\ref{paira'}) becomes 
\begin{equation}
{\partial \over \partial t}(\bar N_{e^\pm}) = -\bar N_{e^\pm}{1\over{\cal
V}}{\partial {\cal V}\over \partial t}+\overline{\sigma
v}{1\over\bar\gamma^2} (\bar N^2_{e^\pm} (T) - \bar N^2_{e^\pm}),
\label{paira}
\end{equation}
where the coordinate number-density of $e^\pm$-pairs in equilibrium is $
\bar N_{e^\pm}(T)\equiv\bar\gamma n_{e^\pm}(T)$ and the coordinate
number-density of $e^\pm$-pairs is $ \bar N_{e^\pm}\equiv\bar\gamma \bar n_{e^\pm}$. For
an infinitesimal expansion of the coordinate volume from ${\cal
V_\circ}$ to ${\cal V}$ in the coordinate time interval $t-t_\circ$,
we can discretize the differential Eq.(\ref{paira}) for numerical computations.

\section{\it Three models of different geometry for the expansion of the PEM-pulse}

We analyze three possible patterns of expansion of the PEM-pulse for those cases where the relationship
between the average relativistic $\bar\gamma$ factor and the radial component of
the four-velocity $U_r(r)$ is determined by the following respective conditions (see Eqs.(\ref{asww},\ref{ga}))

\begin{figure}
   \resizebox{\hsize}{8cm}{\includegraphics{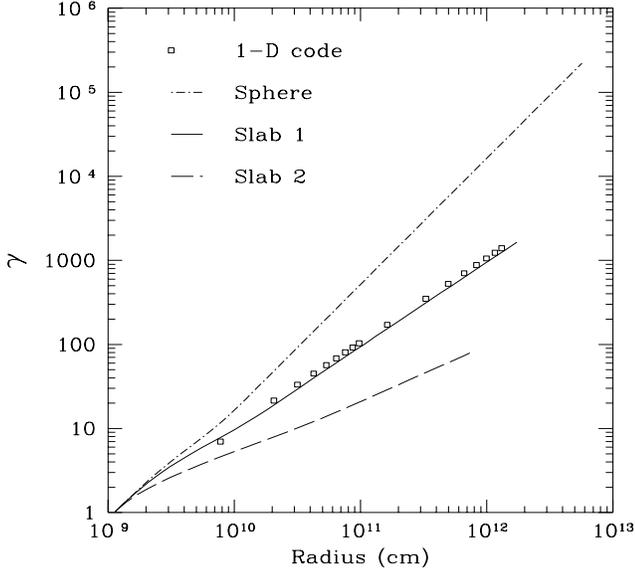}}
      \caption[]{ Lorentz gamma factor $\gamma$ as a function of radius.
Three models for the expansion pattern of the PEM-pulse are
compared with the results of the one dimensional hydrodynamic code for
a $1000 M_\odot$ black hole with charge to mass ratio $\xi=0.1$.  The 1-D
code has an expansion pattern that strongly resembles that of a shell
with constant coordinate thickness.
\label{figshells}}
   \end{figure}

\begin{itemize}

\item   
Spherical model: assuming the radial component of the four-velocity $U_r(r)=U{r\over {\cal R}}$,
where $U$ is the radial component of the four-velocity at the moving outer surface (${r=\cal R}(t)$) of the PEM-pulse,
the $\bar\gamma$-factor and the velocity $V_r$ are
\begin{eqnarray} 
\bar\gamma &=& {3\over 8U^3}\Big[2U(1+U^2)^{3\over2}
- U(1+U^2)^{1\over2}\nonumber\\
&-& \ln(U+\sqrt{1+U^2})\Big],\hskip0.3cm V_r={U_r\over\bar\gamma}~;
\label{sp}
\end{eqnarray}
this distribution expands keeping an uniform density profile, decreasing with time, similar to a portion of a Friedmann Universe.

\item 
Slab 1: assuming $U(r)=U_r={\rm const.}$, the constant width of the
expanding slab ${\cal D}= R_\circ$ in the coordinate frame of the PEM-pulse, $\bar\gamma$  and $V_r$ are
\begin{equation}
\bar\gamma=\sqrt{1+U_r^2},\hskip0.3cm V_r={U_r\over\bar\gamma}~;
\label{sl1}
\end{equation}
this distribution does not need any averaging process.

\item
Slab 2: assuming a constant width $R_2-R_1=R_\circ$ of the expanding slab in the
comoving frame of the PEM-pulse, $\bar\gamma$  and  $V_r$ are
\begin{equation}
\bar\gamma=\sqrt{1+U_r^2(\tilde r)},\hskip0.3cm V_r={U_r\over\bar\gamma},
\label{sp2}
\end{equation}
This distribution needs an averaging procedure and 
 $R_1<\tilde r <R_2$, i.e.~$\tilde r$ is an intermediate radius
in the slab.

\end{itemize}
In these three specific models of relativistic expansion, the
average relativistic gamma factor $\bar\gamma$ 
is related differently to
the radial component of the four-velocity $U(r)$ and the velocity of expansion. As a consequence,
this leads to distinct relationships between the coordinate radius
${\cal R}$ and the coordinate time $t$ in different models. This leads
to the monotonically increasing $\bar\gamma$-factor as a function of
the coordinate radius (or time) having a distinct slope (see
Fig.[\ref{figshells}]). This slope varies from model to model. In
principle, we could have an infinite number of models by defining the
geometry of the expanding fluid.  Thus, to find out which expanding
pattern of PEM-pulses is the most physically realistic, we need to
confront the results of our proposed theoretical simplified models with the
numerical results based on the hydrodynamic
Eqs.(\ref{jay1},\ref{jay2},\ref{jay3}) and see if any one of them is in agreement with these results.

Eqs.(\ref{be},\ref{scale},\ref{result},\ref{paira}) can be used to 
make numerical integrations to relate the quantities $\bar\gamma$,$\bar\epsilon$,
$\bar\rho_B$ and $\bar N_{e^\pm}$ to $\bar\gamma_\circ$, $\bar\epsilon_\circ$,
$\bar\rho_B^\circ$ and $\bar N^\circ_{e^\pm}$ in terms of the ratio ${{\cal
V}_\circ\over {\cal V}}$. We can consider each step of the adiabatic
process of expansion from volume ${\cal V}_\circ$ to volume ${\cal V}$
to be infinitesimal. Given initial values of the quantities
$\bar\gamma_\circ, \bar\epsilon_\circ, \bar\rho_B^\circ $ and
$\bar N^\circ_{e^\pm}$ in the volume ${\cal V}_\circ$ and temperature
$T_\circ$ as well as $\bar N^\circ_{e^\pm}(T_\circ)$, we can obtain the
values of $\bar\gamma$, $\bar\epsilon$ and $\bar\rho_B$ corresponding to the 
volume ${\cal V}$ by using computer to solve
these three dynamical Eqs.(\ref{be},\ref{scale},\ref{result}). With the value of $\bar\gamma$ obtained we 
find the corresponding velocity $V_r$ by Eqs.(\ref{asww},\ref{sp},\ref{sl1},\ref{sp2}) and the coordinate time
$t=t_\circ + \Delta t, \Delta t=\Delta {\cal R}/V_r$, where $t_\circ,t$ are the times corresponding 
to expanding volumes ${\cal V}_\circ, {\cal V}$ respectively and ${\cal R}={\cal R}_\circ +
\Delta {\cal R}$, where $\delta {\cal R}$ is the variation of the
coordinate radius at the surface of the PEM-pulse as its volume changes from volume ${\cal V}_\circ$ to volume ${\cal V}$. Then we find the corresponding 
number of pairs $\bar N_{e^\pm}$ by solving
the rate equation (\ref{paira}). Next we evaluate the fraction ${\bar N_{e^\pm}\over
\bar N_{e^\pm}(T_\circ)}$ and the appropriate Fermi integrals in the right-hand-side of
Eqs.(\ref{eeq},\ref{hat}), so that we can determine the
temperature $T$ of the equilibrium state corresponding to the volume
${\cal V}$ from the proper energy-density through Eqs.(\ref{eeq},\ref{hat}). Then by iterative 
computing steps, ${\cal
V}_\circ,{\cal V}_1,{\cal V}_2,\cdot\cdot\cdot$, we can describe the whole
process of the expanding PEM-pulse. As a result, given initial
values $\bar\gamma=1, {\cal R}_\circ=r_{ds}$, $T_\circ$,
$\bar\rho_B^\circ$, $\bar\epsilon_\circ$ and $\bar N_{e^\pm}= \bar N_{e^\pm}(T_\circ)$ of
a dyadosphere discussed in section 1, we can obtain $\bar\gamma({\cal
R})$ and energy-density $\bar\epsilon({\cal R})$ of the expanding
PEM-pulse in terms of the coordinate radius ${\cal R}$ and time
$t$. The calculations are continued until the plasma becomes optically thin as expressed by the condition (\ref{thin}). In fact, Eqs.(\ref{res},\ref{result}) are equivalent
to
\begin{equation}
{d(\bar\rho\bar\gamma^2{\cal V})\over dt}=0,
\label{tc}
\end{equation}
the conservation of the total energy of the PEM-pulse at each
expanding step, and we check that it is fulfilled at each numerical
iteration step. 

From the conservation of entropy it follows that asymptotically we have
\begin{equation}
      \frac{(V T^3)_{T<mc^2}}{(V T^3)_{T>mc^2}}  =\frac{11}{4}\ ,
\label{reheat}
\end{equation}
exactly for the same reasons and physics scenario discussed in the cosmological framework by Weinberg, see e.g. Eq.~(15.6.37) of \cite{ws}: the present equation follows by substituting  the comoving
volume factor $R^3$ by the comoving volume $V$ in the cited equation. 
The same considerations when
repeated for the conservation of the total energy 
$\bar\rho\bar\gamma V=\bar\rho\bar\gamma^2{\cal V}$
following from Eq.~(\ref{tc}) then lead to
\begin{equation}
      \frac{(V T^4 \bar\gamma)_{T<mc^2}}{(V T^4 \bar\gamma)_{T>mc^2}}  
             =\frac{11}{4}\ .
\end{equation}
The ratio of these last two relations then gives asymptotically 
\begin{equation}
      T_\circ= (T \bar\gamma)_{T>mc^2}= (T \bar\gamma)_{T<mc^2},
\label{rt}
\end{equation}
where $T_\circ$ is the initial average temperature of the dyadosphere at rest, given in Fig.[\ref{fig71}]. 
Eq.(\ref{rt}) explains the approximate constancy of $T \bar\gamma$ shown 
in Fig.[\ref{figallTm}].

%
   \begin{figure}
   \resizebox{\hsize}{8cm}{\includegraphics{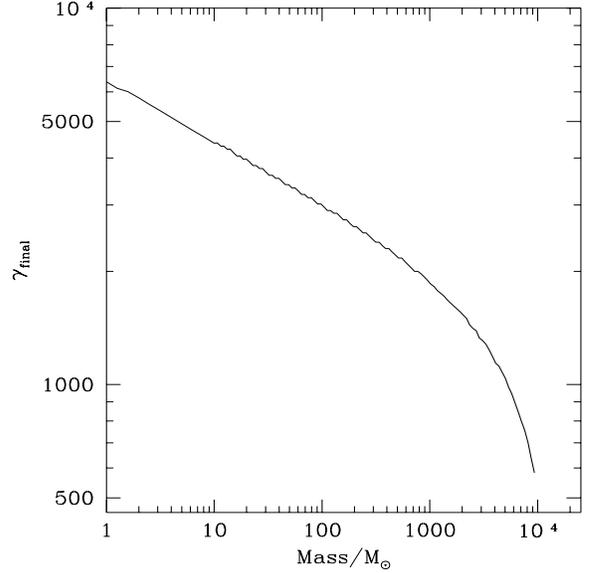}}
      \caption[]{In the expansion model of a shell with
constant coordinate thickness, the decoupling gamma-factor
$\gamma_{\rm final}$ as a function of EMBH masses is plotted with charge to mass ratio
$\xi=0.1$.
\label{dgamma}}
   \end{figure}
%
   \begin{figure}
   \resizebox{\hsize}{8cm}{\includegraphics{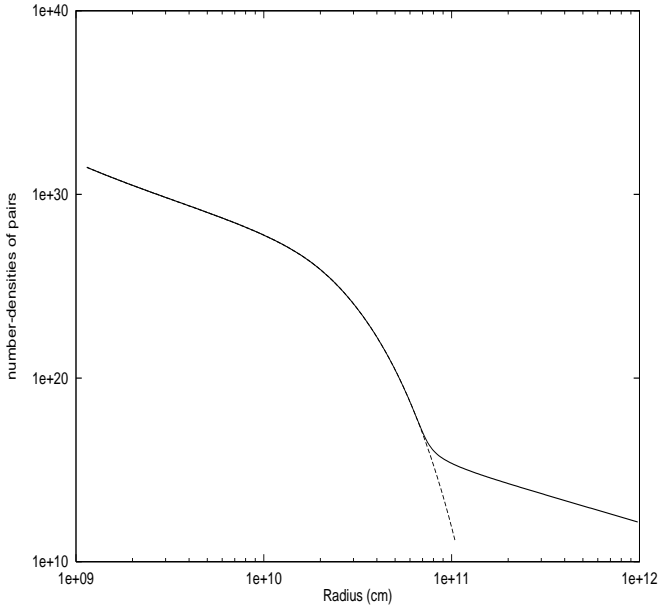}}
      \caption[]{The number-densities ($N_{e^\pm}$/cm$^{-3}$) 
of pairs $n_{e^\pm}$ (solid line)
as obtained from the rate equation (\ref{econtin}), and $n_{e^\pm}(T)$ 
(dashed line) as computed by Fermi-integrals with zero chemical potential, provided the temperature $T$ determined by the equilibrium condition (\ref{eeq}), are plotted for
a $1000 M_\odot$ black hole with charge to mass ratio $\xi=0.1$. For $T< m_ec^2$, the two curves strongly diverge.
\label{figarticlepnpnt}}

   \end{figure}
%
   \begin{figure}
   \resizebox{\hsize}{8cm}{\includegraphics{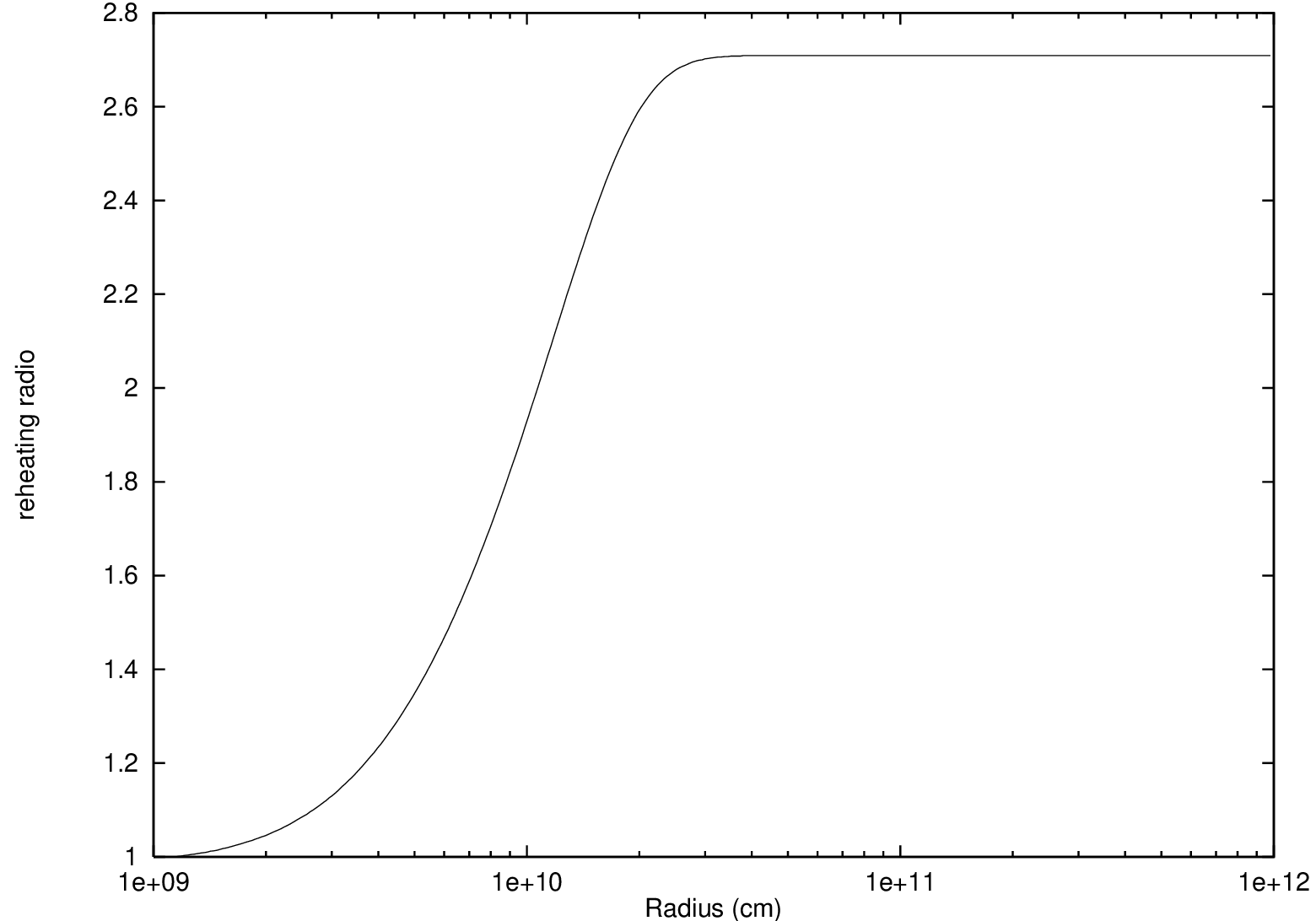}}
      \caption[]{The reheating ratio $T^3V/T^3_\circ V_\circ$ defined by Eq.(\ref{reheat})
is plotted as a function of radius for
a $1000 M_\odot$ black hole with charge to mass ratio $\xi=0.1$. The rate equation (\ref{econtin}) with definition (\ref{hat}) naturally leads to the value ${11\over4}$ after 
$e^+e^-$-annihilation has occurred.
\label{figarticleratio}}
   \end{figure}
%

\section{\it Analysis of the Spectrum and Lightcurve at the Time of Decoupling} \label{specnltcrv}

We are interested in the observed spectrum at the time of decoupling.
To calculate the spectrum, we assume that (i) the plasma fluid of
coupled $e^+e^-$ pairs and photons undergoes adiabatic expansion and
does not emit radiative energy before they decouple (this assumption is relaxed in \cite{pr1} and \cite{prx3}); (ii) the $e^+e^-$ pairs and photons are in
equilibrium at the same temperature $T$ when they decouple.  Thus the photons in
the fluid frame (denoted with a prime: $'$) are described by a Planck
distribution of the form
\begin{equation}
u'_{\epsilon'}(T') \approx { \epsilon'^3  \over exp({\epsilon' \over T'}) - 1},
\end{equation}
where $T'$ is the local temperature in the fluid frame of one thin shell. However, $u_\epsilon / \epsilon^3$ is a relativistic
invariant resulting from the Liouville theorem (see e.g. \cite{eh}).  This implies $\epsilon / T$ is
also a relativistic invariant.  So a Planck distribution in an
emitter's rest-frame with temperature $T'$ will appear Planckian to a
moving observer, but with boosted temperature $T = T'/(\gamma (1 - {v\over c} \cos
\theta))$ where ${v\over c} \cos \theta$ is the component of fluid velocity 
directed toward the observer.  Thus 
\begin{equation}
u_\epsilon (\theta,v,T') \approx { \epsilon^3 \over exp(\gamma (1 - {v\over c} \cos \theta) { \epsilon \over T'}) - 1 }
\label{jay:E:planck}
\end{equation}
gives the observed spectrum of a blackbody with fluid-frame temperature
$T'$ moving at velocity $v$ and angle $\theta$ with respect to the
observer.

We wish to calculate the spectrum from a spherical, relativistically
expanding shell as seen by a distant observer.  We know the velocity
$v$ and proper temperature $T'$ and shell radius $R$, so we
integrate over volume and angle with respect to the observer.  We thus
get the observed number spectrum $N_\epsilon$, per photon energy
$\epsilon$, per steradian, of a relativistically expanding spherical
shell with radius $R$, thickness $dR$ in cm, velocity $v$, Lorentz
factor $\gamma$ and fluid-frame temperature $T'$ to be (in
photons/eV/$4\pi$)
\begin{eqnarray}
N_\epsilon(v,T',R) &\equiv&  \int dV {u_\epsilon
 \over \epsilon}= (5.23 \times 10^{11}) 4\pi R^2 dR {\epsilon T' \over v
\gamma}\nonumber\\
&\cdot& \log \Biggl[ {1 - exp[- \gamma \epsilon (1 + {v\over c})/T' ] \over 1 -
exp[ - \gamma \epsilon (1 - {v\over c})/T' ] } \Biggr],
\label{jay:E:nmax}
\end{eqnarray}
which has a maximum at $\epsilon_{max} \cong 1.39 \gamma T'\ eV$ for
$\gamma \gg 1$.  We may then sum this spectrum over all shells (the
zones in our computer code) of our PEM-pulse to get the total
spectrum.  Since we had assumed the photons are thermal in the comoving frame, our
spectrum has now an high frequency exponential tail, and the spectrum appear as
not thermal.  In Fig.[\ref{denergy}] we see the observed number
spectral peak energy as a function of the EMBH mass with charge to mass ratio $\xi =
0.1$.
   \begin{figure}
   \resizebox{\hsize}{8cm}{\includegraphics{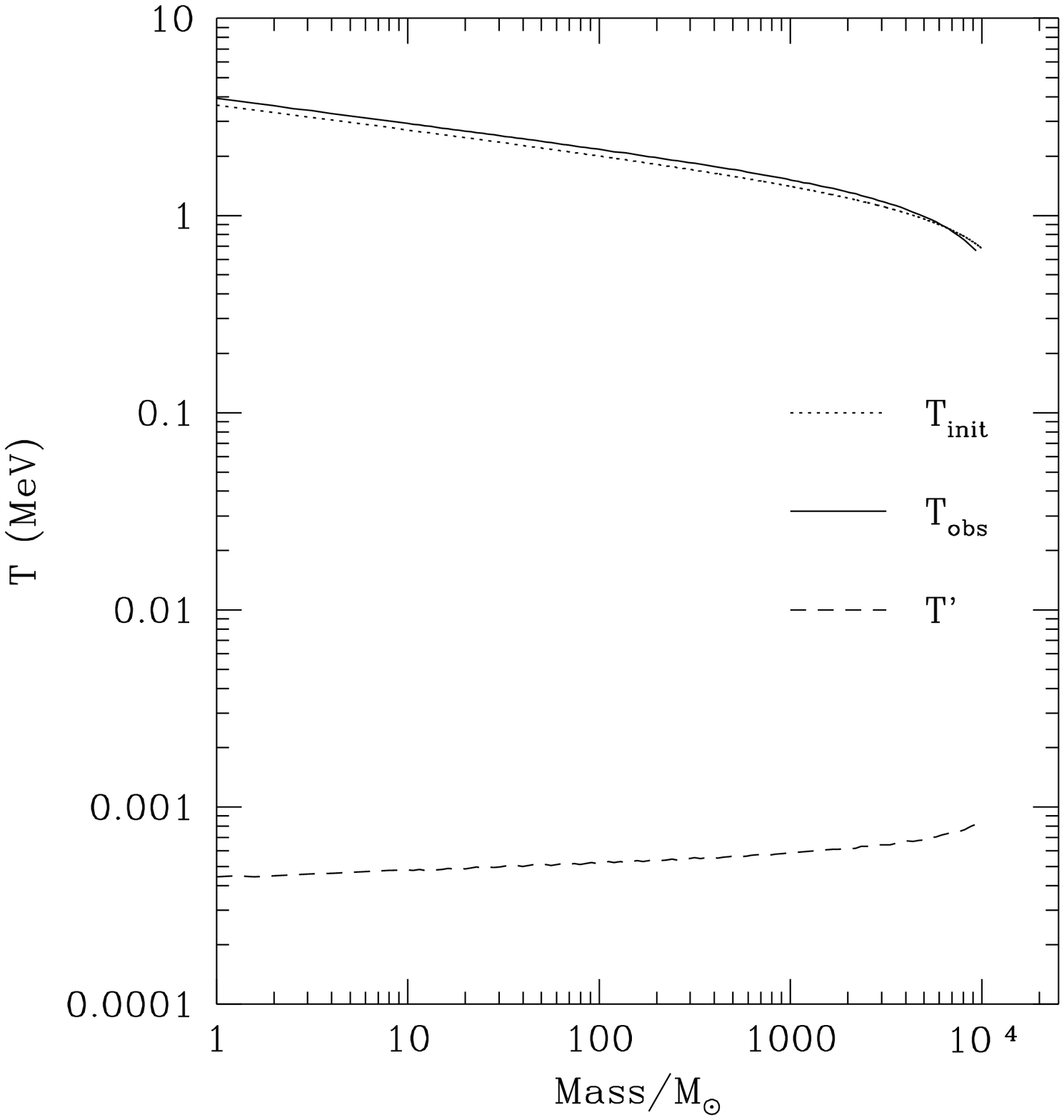}}
      \caption[]{Temperature of the plasma is shown over a
range of masses.  $T_{\rm init}$ is the average initial temperature of the
plasma deposited around the EMBH, $T_{\rm obs} = 1.39 \gamma_{\rm final} T'$
is the observed temperature of the plasma at decoupling while $T'$ is
the comoving temperature at decoupling.  Notice that $\gamma T'\simeq T_{\rm init}$  as expected from Eq.(\ref{rt}).
 \label{figallTm}}
   \end{figure}
%

   \begin{figure}
   \resizebox{\hsize}{8cm}{\includegraphics{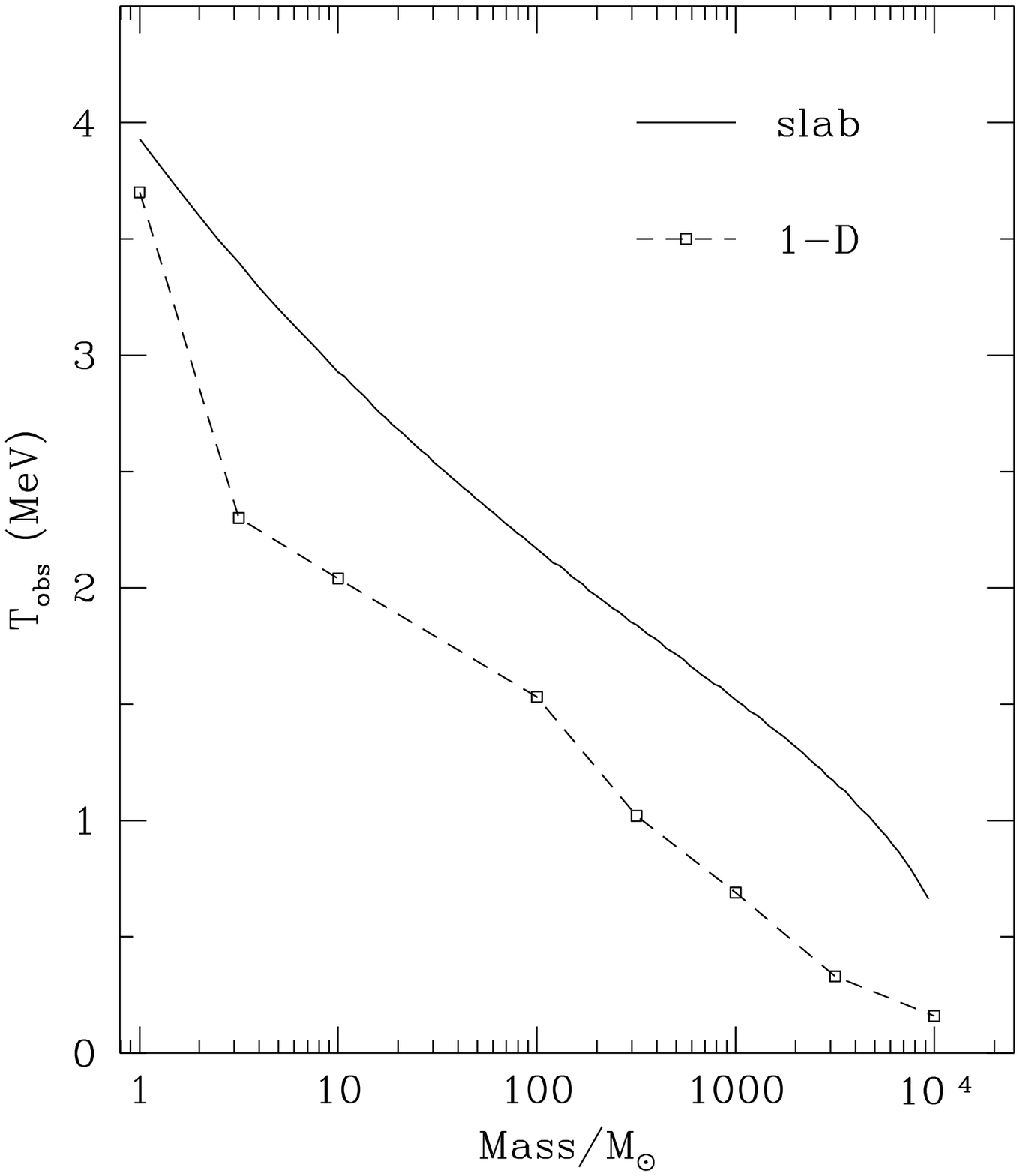}}
      \caption[]{The peak of the observed number spectrum as a
function of the EMBH mass is plotted with charge to mass ratio $\xi=0.1$.  The observed
peak temperature $T_{\rm obs}$ of the number spectrum is calculated from
Eq.(\ref{jay:E:nmax}).  The temperature for the slab of
constant coordinate thickness is given by $T_{\rm obs} = 1.39
\gamma_{\rm final} T'$; the peak of Eq.(\ref{jay:E:nmax}).
\label{denergy}}
   \end{figure}
%

To obtain the observed light curve $\varepsilon(t)$ we again
decompose the spherical PEM-pulse into concentric shells and consider two
effects.  First is the relative arrival time of the first light from
each shell: light from outer shells will be observed before light from
inner shells.  Second is the shape of the light curve from a single
shell.

Emission from the moving PEM-pulse is beamed along the direction of travel
within an angle $\theta \sim 1/\gamma$.  The surface of simultaneity
of a relativistically expanding spherical shell as seen by an observer
is an ellipsoid (see e.g. \cite{jay:fenimore}).  The time of intersection of an
expanding ellipse and a fixed shell of radius R as a function of
$\theta$ (i.e. the time at which emission from this intersection
circle is received) is
\begin{equation}
t = \frac{R}{v}(1 - {v\over c} \cos \theta) = (1 - \cos \theta) R + \frac{R}{v}(1 - {v\over c})~.
\end{equation}
We find that, integrating our boosted Planck distribution of photons
Eq.(\ref{jay:E:planck}) over frequency, a relativistically
expanding shell of radius R and thickness dR will have a time profile
(in ergs/second/$4\pi$)
\begin{equation}
\varepsilon(\tau,v,T',R) = \frac{a}{2} \biggl( \frac{T'}{\tau}
\biggr)^4 \frac{cR^2 dR}{v \gamma^4}
\label{jay:E:ltcurve}
\end{equation}
for $(1-{v\over c}) \leq \tau \leq (1 + {v\over c})$, where $\tau \equiv {vt \over R}$
and $a$ is the radiation constant.  Thus the larger the radius $R$ of
the expanding shell at the moment of emission, the broader the $1 /
\tau^4$ tail on the light curve.  The final light curve is then
constructed by summing the signal from all shells.

\begin{figure}
   \resizebox{\hsize}{8cm}{\includegraphics{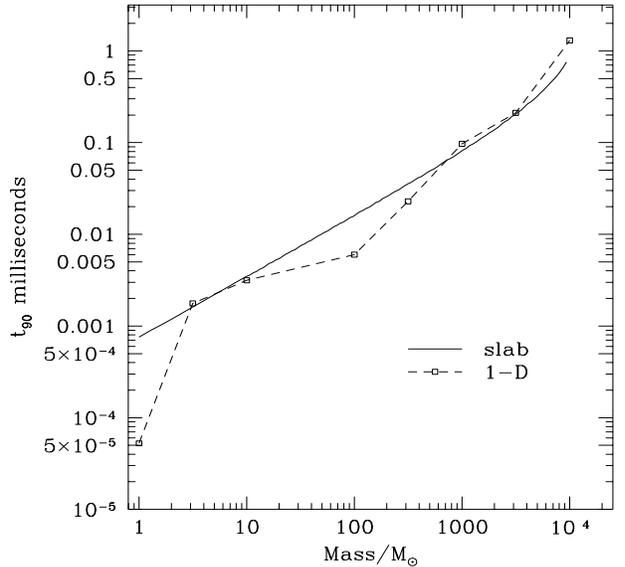}}
      \caption[]{The duration of the emission at decoupling is
represented by $t_{90}$, the time over which 90\% of the emission is
received, plotted over a range of black hole masses.  The 1-D
hydrodynamic $t_{90}$ is calculated from Eq.(\ref{jay:E:ltcurve}) and
the slab $t_{90}$ is calculated from Eq.(\ref{jay:E:ltcurve2}).  Note
that the 1-D and slab models deviate for black hole mass $M = 1M_\odot$.  This
is because the energy density Eq.(\ref{jayet}) deposited in the 1-D
computations is not well approximated by a constant over the volume as
in the slab model for small black hole mass.
\label{figt90}}
   \end{figure}

A useful quantity which may be compared with observations is $t_{90}$;
the time over which 90\% of the emission is received.  The $t_{90}$
duration for a single shell can be calculated in the relativistic
limit ($\gamma \gg 1$) by defining $\varepsilon(\tau) \equiv
\varepsilon_0 \tau^{-4}$ and setting
\begin{equation}
\int_{1-{v\over c}}^{1-{v\over c}+\tau_{90}} \varepsilon (\tau) d\tau = 0.9 \int_{1-{v\over c}}^{1+{v\over c}} 
\varepsilon (\tau) d\tau,
\end{equation}
which gives $\tau_{90} \approx 4.3/\gamma^2$.  So the emission
duration in seconds is
\begin{equation}
t_{90} \approx \tau_{90} \frac{R}{c} = 4.3 \frac{R}{\gamma^2
c} 
\label{jay:E:ltcurve2}
\end{equation}
for $\gamma \gg 1$.  Burst durations are shown in Fig.[\ref{figt90}]
for a range of EMBH masses with charge to mass ratio $\xi = 0.1$.
%

\section{\it Results and Discussion}

Preliminary results have been presented in Ruffini (1999a), 
Ruffini et al.(1999) and Ruffini (1999b).
In the results presented here we consider a range of black hole masses from $1$ to $10^4 M_\odot$ and we assume a charge of $Q = 0.1 Q_{max}$ $(\xi = 0.1)$.
Over this mass range, and with this charge, a large amount of energy
($\sim 10^{52} - 10^{54}$ ergs) is available to be deposited within the dyadosphere, in the
form of $e^+e^-$ pairs,  due to the process of vacuum polarization (Fig.[\ref{figetot}]). In Fig.[\ref{fig71}] correspondingly the average energy of the created pairs are given as a function of the EMBH mass. These energies and spectral distributions are in the range of interest for gamma-ray bursts.

We have  analyzed the evolution of this PEM pulse by a simple relativistic treatment in flat space for three different cases: 1) an expansion with a spatial component of the four velocity proportional to the distance, 2) an expansion with a width of the PEM pulse constant in the coordinate space and 3) an expansion with a width of the PEM pulse constant in the comoving frame. We have then compared and contrasted these results with the general relativistic hydrodynamic calculations. In Fig.[\ref{figshells}] we see a comparison of the Lorentz factor of
the expanding fluid as a function of radius for all of the models.
We can see that the one-dimensional code matches the expansion
pattern of a shell of constant coordinate thickness. The
astrophysically unprecedented large Lorentz factors are reproduced in Fig.[\ref{dgamma}] as a
function of the EMBH mass.

In Fig.[\ref{figarticlepnpnt}] and Figs.[\ref{figarticleratio},\ref{figallTm}] we describe the approach to transparency, as given by Eq.(\ref{thin}). In Fig.[\ref{figarticlepnpnt}] we plot the number-densities
of pairs $n_{e^\pm}$ given by the rate equation (\ref{econtin}) and 
$n_{e^\pm}(T)$ computed from a Fermi-integral with zero chemical potential with
the temperature $T$ determined by the equilibrium condition (\ref{eeq}). 
It clearly indicates that the pairs $e^\pm$ fall out of equilibrium as 
the temperature drops below the
threshold of $e^\pm$-pair annihilation. As a consequence of pair 
$e^\pm$-annihilation, the crossover of this reheating process is shown in 
Fig.[\ref{figarticleratio}]. Fig.[\ref{figallTm}] 
illustrates the extreme
relativistic nature of the PEM pulse expansion: at decoupling, the
local comoving plasma temperature is $< 1$ keV, but is boosted by
$\gamma_{\rm final}$ to $\sim 1$ MeV and its relation to the initial energy of the PEM pulse in the dyadosphere is presented.

The spectrum and light-curve of the emitted pulse of radiation are analyzed as described in Section \ref{specnltcrv}. We see the results
for both the hydrodynamic and slab calculations compared in
Figs.[\ref{denergy}] \& [\ref{figt90}].  The 1-D hydrodynamic code
spectral peak and burst duration $t_{90}$ are consistently somewhat lower than
those of the slab calculations due to gravitational and hydrodynamical
effects.  The peak photon energy of the photon number spectrum is in
the range of $100 - 1000$ keV, which is consistent with observed
gamma-ray bursts.  The duration of the burst is characterized by the
$t_{90}$ time, the time taken to receive $90\%$ of the radiation.
Over the mass range analyzed, $t_{90}$ is of order a millisecond.
This is shorter than observed gamma-ray bursts ($\sim 1-10$ seconds, see e.g.~\cite{ta}),
but the collision of the PEM pulse with the remnant left over by the process of gravitational collapse produces
longer emission times, see e.g. \cite{rssw1}, \cite{pr1}
and \cite{prx3}.

We have demonstrated that the concept of an EMBH in vacuum can produce
a gamma-ray burst having many of the general characteristics of
observed bursts validating the scenario presented in \cite{rr}. This simplified model merits further detailed study of the collapse of highly magnetized
stars in galaxies ($M\sim 10M_\odot$) and gravitational collapse process occurring in galactic nuclei ($M\sim 10^4-10^5M_\odot$).  In particular, it is important to identify
how, during the process of gravitational collapse to an EMBH, the
breakdown of the magneto-hydrodynamic conditions leads to the
formation of the dyadosphere.  Future work will explore some
additional important aspects of the PEM pulse including (i) the presence of
baryonic matter left in the remnant around the EMBH during the process of gravitational collapse and its effect on the relativistic
expansion (\cite{rssw1}), (ii) the
radiation emitted prior to decoupling and its effect on the dynamics of the PEM pulse (\cite{pr1}
and \cite{prx3}), and (iii) the emission in the latest stages of evolution of the PEM pulse and its comparison with the observed bursts and afterglows in (\cite{pr1} and \cite{prx3}).\\

J.\ D.\ Salmonson and J.\ R.\ Wilson were supported under the auspices of
the U.S. Department of Energy by the Lawrence Livermore National
Laboratory under contract W-7405-ENG-48.  J.\ R.\ Wilson has additional
support from NSF grant No.\ PHY-9401636.

It is a pleasure to thank the referee and R.\ Jantzen for helpful suggestions in the wording of the manuscript.


\begin{thebibliography}{}

\bibitem[Baring \& Harding 1998]{sgr2}
Baring, M.\ G.\, Harding, A.\ K. 1998, ApJ 507, L55.

\bibitem[Bloom {\it et al.} 1999]{bloom} Bloom, J.\ S.\, Odewahn, S.\ C.\,
Djorgovski, S.\ G.\ {\it et al.} 1999, ApJ 518, L1. 

\bibitem[Christodoulou \& Ruffini 1971]{rc}Christodoulou D. , Ruffini
R. 1971, {\it Phys.~Rev.} {\bf D4}, 3552.

\bibitem[Damour \& Ruffini 1975]{dr}
Damour T.~, Ruffini R. 1975, {\it Phys.~Rev.~Lett.} {\bf 35} (1975) 463.

\bibitem[Damour {\it et al.} 1978]{dhrw}
Damour T.~, Hanni R.~S.~, Ruffini R., Wilson J.~R. 1978, {\it Phys.~Rev.} {\bf D17}, 1518.

\bibitem[Damour \& Ruffini 1999]{tdrr}
Damour T.~, Ruffini R. 1999, paper in preparation.

\bibitem[Ehlers  1971]{eh}
Ehlers J.~1971, Proceedings of cours 47 of the International School of Physics: Enrico Fermi, ed. Sachs, R.K.~, Academic Press, New York.

\bibitem[Frontera \& Piro 1999]{fp}Frontera F.~, Piro L.~eds., Proc. of \lq\lq Gamma-ray bursts in the afterglow era", Rome Nov.3-6 1998, to be published in Astron. \& Astrophys. Suppl.~Series, 1999 and references therein

\bibitem[Fenimore {\it et al.} 1996]{jay:fenimore}Fenimore, E.\ E.\, Madras C.\, Nayakshin S.\
1996, ApJ 473, 998

\bibitem[Heisenberg \& Euler 1931]{he}
Heisenberg W.\ , Euler H. 1931, Zeits. Phys. 69, 742

\bibitem[Kouveliotou {\it et al.} 1998]{sgr1} 
Kouveliotou, C.\, Dieters, S.\, Strohmayer, T.\ {\it et al.} 1998, Nat. 393, 35

\bibitem[Preparata, Ruffini \& Xue 1998a]{prxa}Preparata, G.\, Ruffini,
R.\, Xue, S.-S. 1998a, submitted to {\it Phys.~Rev.~Lett.}

\bibitem[Preparata, Ruffini \& Xue 1998b]{prxb}Preparata G.\, Ruffini,
R.\, Xue, S.-S. 1998b, A\&A 338, L87.

\bibitem[Preparata \& Ruffini 1999]{pr1} Preparata G., Ruffini,
R. 1999,  submitted for publication.

\bibitem[Preparata, Ruffini \& Xue 1999]{prx3} Preparata G.\, Ruffini,
R.\, Xue, S.-S. 1999, paper in preparation.

\bibitem[Ruffini \& Wilson 1975]{wr}
Ruffini R.\, Wilson J.~R. 1975, {\it Phys.~Rev.} {\bf D12}, 2959. 

\bibitem[Ruffini 1998]{rr}
Ruffini R. 1998, in \lq\lq Black Holes and High Energy Astrophysics". Proceedings 
of the 49th Yamada Conference Ed. H. Sato and N. Sugiyama. Universal Ac. 
Press Tokyo 1998.

\bibitem[Ruffini 1999a]{rro}
Ruffini R. 1999a, \lq\lq The dyadosphere of black holes and gamma-ray bursts", 
Proc. of \lq\lq Gamma-ray bursts in the afterglow era", Rome Nov.3-6 1998, to be published in Astron. \& Astrophys. Suppl.~Series, 1999, Ed.~by Piro L. \& Frontera F.

\bibitem[Ruffini {\it et al.} 1999a]{rr4}
Ruffini R.\, Salmonson J.\, Wilson J. R.\, Xue S.-S. 1999a, \lq\lq On structure and temporal evolution of pair and electromagnetic pulse of an electromagnetic black hole", Proc. of \lq\lq Gamma-ray bursts in the afterglow era", Rome Nov.3-6 1998, to be published in Astron. \& Astrophys. Suppl.~Series, 1999, Ed.~by Piro L. \& Frontera F.

\bibitem[Ruffini {\it et al.} 1999b]{rssw1}
Ruffini R.\, Salmonson J.\, Wilson J. R.\, Xue S.-S. 1999b, paper in preparation.

\bibitem[Ruffini 1999b]{rrp}
Ruffini R. 1999b, \lq\lq The Dyadosphere of black holes and Gamma Ray Bursts", Proc. of 19th Texas Symposium, Paris Dec.~1998.

\bibitem[Schwinger 1951]{sw}
Schwinger J. 1951, {\it Phys.~Rev.} {\bf 82}, 664.

\bibitem[Tavani 1998]{ta}
Tavani M. 1998, ApJ 497, L21.

\bibitem[Weinberg 1972]{ws} Weinberg S. 1972, {\em Gravitation and Cosmology}
(USA: John Wiley \& Sons, Inc.)

\bibitem[Wilson 1975]{jrw1} Wilson J.\ R. 1975, Annals of the New York
Academy of Sciences 262, 123

\bibitem[Wilson 1977]{jrw2} Wilson J.\ R. 1977, in {\it Proc. of the
First Marcel Grossmann Meeting on General Relativity}, ed. R. Ruffuni
(North-Holland Pub. Amsterdam) 1977, p. 393.

\bibitem[Wilson, Salmonson \& Mathews 1997]{jay:wsm97}
Wilson, J.\ R.\, Salmonson, J.\ D., Mathews, G.\ J. 1997,
in {\it Gamma-Ray Bursts: 4th Huntsville Symposium}, ed. C.\ A.\
Meegan, R.\ D.\ Preece, T.\ M.\ Koshut (American Institute of Physics)

\bibitem[Wilson, Salmonson \& Mathews 1998]{jay:wsm98}
Wilson, J.\ R., Salmonson, J.\ D., Mathews, G.\ J. 1998, in {\it 2nd Oak Ridge
Symposium on Atomic and Nuclear Astrophysics} (IOP Publishing Ltd) 

\end{thebibliography}
\end{document}